\documentclass[twocolumn]{aastex62}

\usepackage{natbib}
\usepackage[normalem]{ulem} 

\received{datum, 2019}
\revised{datum, 2019}
\accepted{datum, 2019}
\submitjournal{ApJ}

\shorttitle{Damping of Propagating Kink Waves}
\shortauthors{Tiwari et al.}

\begin{document}

\title{Damping of Propagating Kink Waves in the Solar Corona}

\correspondingauthor{Ajay K. Tiwari}
\email{ajay.tiwari@northumbria.ac.uk}
\author[0000-0001-6021-8712]{Ajay K. Tiwari}
\affil{Northumbria University \\
Newcastle upon Tyne, NE1 8ST, UK}

\author[0000-0001-5678-9002]{Richard J. Morton}
\affiliation{Northumbria University \\
Newcastle upon Tyne, NE1 8ST, UK}
\author[0000-0001-8954-4183]{St{\'e}phane R{\'e}gnier}
\affiliation{Northumbria University \\
Newcastle upon Tyne, NE1 8ST, UK}
\author[0000-0002-7863-624X]{James A. McLaughlin}
\affiliation{Northumbria University \\
Newcastle upon Tyne, NE1 8ST, UK}



\begin{abstract}
Alfv\'enic waves have gained renewed interest since the existence of ubiquitous propagating kink waves were discovered in the corona. {It has long been suggested that Alfv\'enic} waves play an important role in coronal heating and the acceleration of the solar wind. To this effect, it is imperative to understand the mechanisms that enable their energy to be transferred to the plasma. Mode conversion via resonant absorption is believed to be one of the main mechanisms for kink wave damping, and is considered to play a key role in the process of energy transfer. This study examines the damping of propagating kink waves in quiescent coronal loops using the Coronal Multi-channel Polarimeter (CoMP). A coherence-based method is used to track the Doppler velocity signal of the waves, enabling us to investigate the spatial evolution of velocity perturbations. The power ratio of outward to inward propagating waves is used to estimate the associated damping lengths and quality factors. To enable accurate estimates of these quantities, {we provide the first derivation of a likelihood function suitable for fitting models to the ratio of two power spectra obtained from discrete Fourier transforms. Maximum likelihood estimation is used to fit an exponential damping model to the observed variation in power ratio as a function of frequency.} We confirm earlier indications that propagating kink waves are undergoing frequency dependent damping. Additionally, we find that the rate of damping decreases, or equivalently the damping length increases, for longer coronal loops that reach higher in the corona. 
\end{abstract}



\section{Introduction}\label{s:intro}

{Magnetohydrodynamic (MHD)} waves are a common phenomena in the solar corona and a plethora of different wave modes have been observed in recent years as instrumentation has become increasingly sophisticated, offering higher spatial and temporal resolutions. There have been several reviews extensively discussing the waves observed in the solar corona 
\citep[e.g.][]{naka2003,asch2004,naka2005,dipu2007,2016review,2016wang}. 

Of the different MHD wave modes, Alfv\'enic waves are {considered} one of 
the main candidates for explaining the raised temperature in the corona. Here the term {Alfv\'enic} refers to MHD wave modes that have properties similar to the idealized Alfv\'en wave in a homogeneous plasma, namely that
they are transverse, with high incompressibility and magnetic tension is the dominant restoring force \citep{marcel2009}. The first detection of transverse wave modes occurred after the launch of the Transition Region and Coronal Explorer (TRACE, see \citet{TRACE1999}), observing the presence of standing kink waves that were 
{excited} sporadically in coronal loops after nearby flaring activity \citep[e.g.][]{asch1999,naka1999}. 
The kink waves are typically found to be rapidly damped, with periods of $\approx4$ minutes and damping time of 
$\approx14$ minutes \citep[e.g.][]{ASCetal2002,verwichte2013,GODetal2016}. The damping of these waves was 
suggested to be due to resonant absorption, a phenomenon present in inhomogeneous plasmas that converts the 
energy in transverse motions to azimuthal motion via resonant coupling 
\citep[e.g.][]{ruderman2002,gossens2002,asch2003}. In the presence of structuring in the direction 
perpendicular to the magnetic field (i.e. the loop plasma is considered denser than the ambient plasma), 
transverse motions generate an intrinsic coupling between the kink (transverse) and Alfv\'en (azimuthal, $m = 1$) modes.
The coupling takes place in a dissipative layer at the loop boundary, located at the resonant point where the 
kink frequency, which lies between the internal and external Alfv\'en frequencies, matches the local Alfv\'en 
wave frequency \citep[e.g.][]{asch2003,goossens2006,antolin2015}. 

In contrast, the propagating kink wave mode was only identified a decade ago 
\citep[e.g.][]{tom2007,MCIetal2011,THUetal2014,MORetal2015} and it is found to be ubiquitous throughout the corona. The excitation mechanism(s) of the propagating kink waves are still not evident. It is believed that horizontal motions of magnetic elements in the photosphere are a key driver of relatively high-frequency ($f>1$~mHz) Alfv\'enic modes \citep[e.g.][]{cravan2005,VANBetal2011}, although the observations from CoMP \citep{tom2007,MORetal2016,MORetal2019} appear to suggest the observed Alfv\'enic waves are, at least partially, excited by \textit{p}-modes \citep{cally2017}. 

\medskip
Given their ubiquity, there has been relatively few observational studies of the propagating kink waves. \cite{tom2009} noted that the propagating kink modes observed in a quiescent coronal loop were damped, with \cite{terradas2010} and \cite{verth2010} suggesting that resonant absorption provides a reasonable description of the observed damping. 
The role of resonant damping of propagating transverse waves is substantiated in 3D, full MHD numerical simulations \citep[e.g.][]{PASetal2010,PASetal2012, MAGVAN2016,PAGDEM2017, PAGDEM2019}. 

\cite{terradas2010} provided an analytic investigation into the role of resonant absorption in the damping of 
propagating kink waves along magnetic flux tubes. We introduce here a number of equations from this theoretical
modeling that we will use in the following study. The assumptions of the model result in an exponentially 
damped profile for the wave, however, there is a suggestion that the kink waves may undergo an initial phase of
Gaussian damping \citep{pascoe2016}. The waves that can be observed by CoMP fall under the long wavelength 
regime, thus the damping length, $L_{D}$, for the propagating kink waves is given by
\begin{equation}
L_{D}=\upsilon_{ph} \xi \frac{1}{f},
\label{eqn:damp_len}
\end{equation}
where $\upsilon_{ph}$ is the phase velocity and $f$ is the frequency. $\xi$ is the equilibrium parameter that takes into account the physical conditions of the flux tube and is given by 
\begin{equation}
\xi=\alpha \frac{1}{m} \frac{R}{\ell} \frac{\rho_{i}+\rho_{e}}{\rho_{i}-\rho_{e}}, \hspace{0.5cm}
m>0
\label{eqn:eqparm}
\end{equation}
where $m$ is the mode number, $R$ is loop radius, $\ell$ is the thickness of the density inhomogeneity layer, $\rho_{i}$ and
$\rho_{e}$ are internal and external densities of the magnetic flux tube, respectively, and $\alpha$ is a
constant whose value describes the gradient in density across the resonant layer. The equilibrium parameter 
is a dimensionless quantity, and can be written in terms of the wavelength $\lambda$, 
\begin{equation}
\xi = \frac{L_{D}}{\lambda}, 
\end{equation}
hence $\xi$ can also be interpreted as the quality factor of the wave damping.

In a companion paper, \citet{verth2010} use the CoMP observations of \citet{tom2009} to estimate the equilibrium parameter. Following \citet{verth2010}, we focus our attention on half of a coronal loop and assume the kink waves at the coronal footpoint of the segment (driven by a non-specific mechanism) have a certain power spectrum, $P_{out}(f)$, where the subscript \textit{out} refers to the fact they are outwardly propagating along this segment. They propagate along the loop and are damped to some degree when they reach the loop apex, at a distance $L$ from the coronal base (considered the half loop length). Waves are also excited at the other footpoint, likely with a similar power spectrum,  $P_{in}(f)$, and we denote these as inwardly propagating. By the time they have reached the apex they have already traveled a distance $L$, and are damped further as they propagate down towards the first footpoint. Assuming exponential damping, we can calculate the average power spectra of the outward and inward waves along the half-loop segment of interest, and the ratio of the two integrated power spectra is found to be 

 \begin{equation}
 	\langle{P(f)}\rangle_{ratio}=\frac{P_{out}(f)}{P_{in}(f)} \exp\left(\frac{2L}{\upsilon_{ph}\xi}f\right). \label{eqn:power_rat}
 \end{equation} 	

This expression will provide the underlying model for the following analysis of propagating kink waves. Utilizing data from CoMP enables us to provide estimates for: the values of the inward and outward power spectra as a function of frequency, the half loop length and the propagation speed of the waves.
This, in combination with Equation~(\ref{eqn:power_rat}), provides us with a means to measure the quality factor ($\xi$) if $\langle{P(f)}\rangle_{ratio}$ is known.

\begin{figure*}[!ht]
\centering
\includegraphics[scale=0.33, trim={10cm 0cm 4cm 0cm},clip,height=0.4\textwidth]{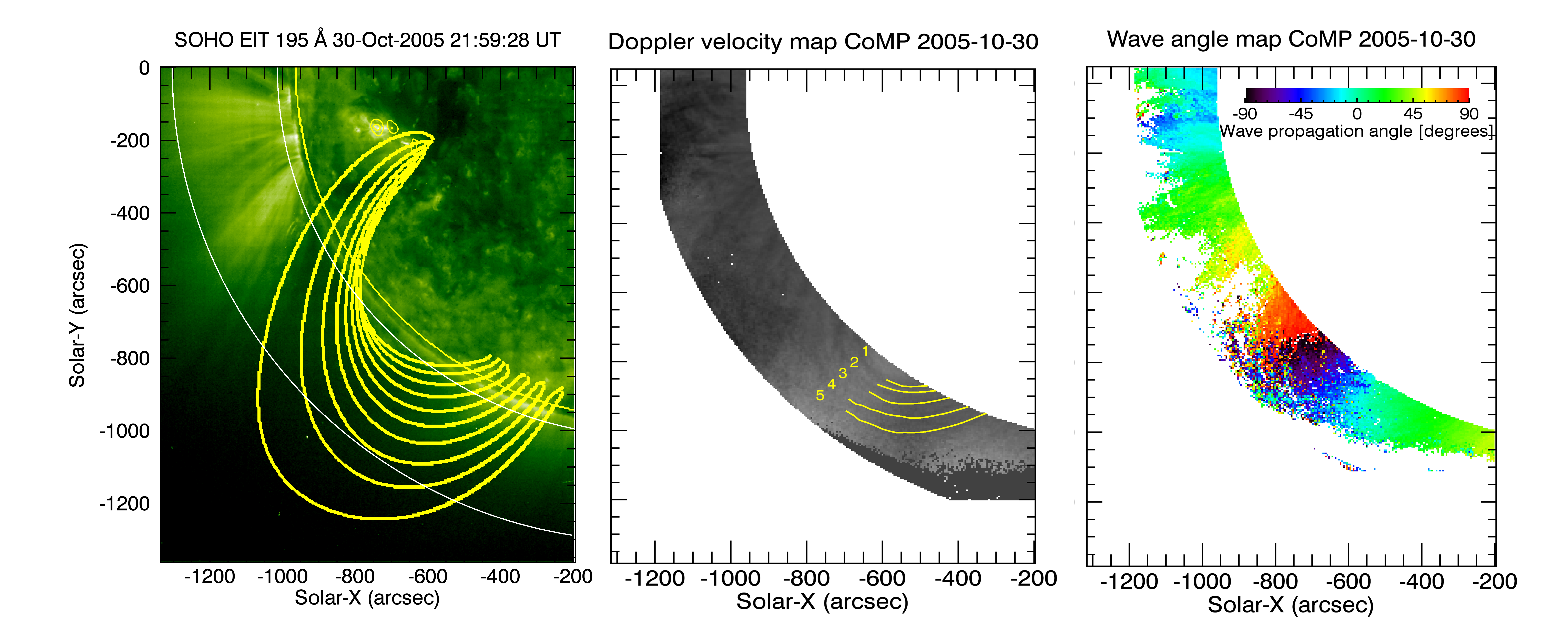}
\caption{\textit{Left:} The PFSS extrapolated magnetic field lines. The field lines were then plotted over the corresponding SOHO/EIT 195 {\AA} image. The lines in white corresponds to the field of view for CoMP for comparison. \textit{Center:} A sample Doppler velocity image is displayed with the over-plotted wave propagation tracks selected for analysis (yellow). \textit{Right:} The wave angle map obtained from the coherence wave tracking method described in Section \ref{wave-angle}. \label{fig:dopplwave}}
\end{figure*}

\bigskip

In order to accurately measure the quality factor, the model in Equation (\ref{eqn:power_rat}) will have to be fit to the power ratio as a function of frequency, as undertaken in \cite{verth2010}. \citet{verth2010} used the least-squares method to achieve this, which assumes that the individual ordinates of the power spectra ratio are normally distributed about their true value. The statistics of the power spectrum obtained via the discrete Fourier transform (DFT) is well studied \citep[ e.g.][]{JENWAT1969,GRO1975,GEWPOR1983,APP2003,vaughan2005}; where the ordinates are known to be distributed about the true values as $\chi^2$ with $\nu$ degrees of freedom ($\nu$ depends on the number of power spectra averaged). Hence, it should not be expected that ordinates from the ratio of two power spectra are normally distributed. In the following we derive for the first time, to the best of our knowledge, the appropriate distribution for the ratio of two $\chi^2_\nu$ distributed power spectra, demonstrating that the assumption of normality, and therefore the utilization of least-squares method, is inappropriate.

\bigskip 
The paper is structured as follows: in Section~\ref{s:data} we provide details
of the data used. The method of analysis is described in Section~\ref{analysis}, where we provide a discussion on the statistics of a power spectrum obtained from DFT and derive the applicable likelihood function required for the maximum likelihood estimation of model parameters from the measured power ratio of damped propagating kink waves. In Section~\ref{s:result}, a discussion of the main findings are given and a conclusion is presented in Section~\ref{s:conclusion}. The tests to validity of the likelihood function we derive is discussed in detail in Appendix \ref{s:app}. We also present a modified model for future analysis of damping in Appendix \ref{s:app3}.

\bigskip
\medskip
\bigskip
\section{Observation}\label{s:data}

The data were obtained using the Coronal Multi-channel Polarimeter (CoMP) \citep{tom2007,tomczyk2008}. CoMP is a combination polarimeter and narrow-band tunable filter that can measure the complete polarization state in the vicinity of the 10747~{\AA} and 10798~{\AA} Fe {\sc{XIII}} coronal emission lines. The data were taken on 30 October 2005, with a temporal cadence of 29~s, and a pixel size of 4$\farcs$5. We focus on the spectroscopic data from the 10747~{\AA} Fe {\sc{XIII}} line, which has been previously used by \citet{tom2007} and \citet{verth2010}. The full details of data acquisition and reduction of the data are described in \citet{tom2007}. The data set consists of Doppler velocity images of the corona between 1.05 $R_{\sun}$ and 1.35 $R_{\sun}$.  An example image is shown in the center panel of Figure \ref{fig:dopplwave}. Here, we will focus our attention on the same off-limb quiescent coronal loops studied previously in \citet{tom2009} and \citet{verth2010}. To provide context images and magnetic field measurements, data from the Solar and Heliospheric Observatory (SOHO) \citep{sohopaper} will also be utilized. Data from the Extreme Imaging Telescope (EIT) \citep{eitpaper} provides a context to the loops observed using CoMP. The background image in the left panel of Figure \ref{fig:dopplwave} is obtained from EIT 195~{\AA} passband. Line-of-sight (LOS) magnetograms from the Michelson Doppler Imager (MDI) instrument \citep{MDIpaper} provide information on the photospheric magnetic field for potential field extrapolations.

\section{Analysis} \label{analysis}
 
\subsection{Extrapolation of the loops}\label{pfss}
\begin{figure}[!t]
\centering
\includegraphics[scale=0.70,clip=true,viewport=50 80 390 290] {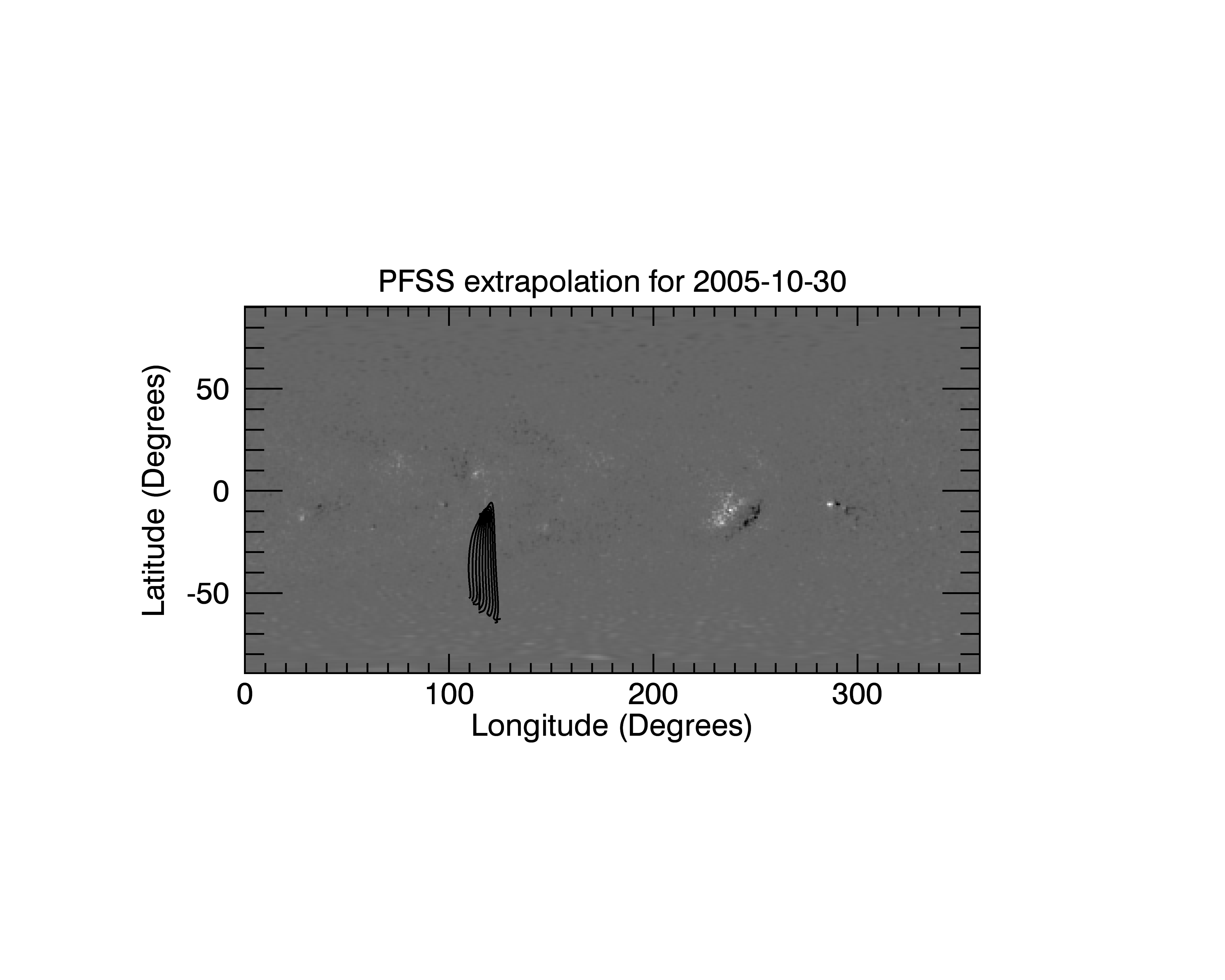}
\caption{Extrapolated field lines for the loops we are interested in, tracked later using wave angles. The field lines are projected on the magnetogram, along the edge-on view.}
\label{fig:pfss}
\end{figure}

Before examining the velocity signals from CoMP, it is beneficial to understand the geometry of the loop system that will be considered. To provide some insight, the potential field source surface (PFSS - \citealp{derosapfss}) extrapolation package available in SolarSoft is used
to provide an indication of the local magnetic field structure in the corona (Figure~\ref{fig:dopplwave}: left). In order to determine the validity of the obtained field extrapolations, several attempts to generate extrapolations in the neighborhood of the footpoints are undertaken, and we find that the given PFSS loops are indeed unique. Further extrapolations were undertaken examining the solution for constant latitudinal points in order to ascertain that the loops we obtained from the initial extrapolations are the best representation for the observed CoMP loops. The extrapolated field lines obtained after these initial checks are shown (Figure~\ref{fig:dopplwave}: left) and visually represent the coronal structures well. There is also close agreement with the direction of wave propagation determined from CoMP, which is believed to follow the magnetic field lines (Figure~\ref{fig:dopplwave} center and right panels, see Section~\ref{wave-angle} for further details). The projection of the field lines onto the magnetogram is shown in Figure~\ref{fig:pfss}. 

\subsection{Determining Wave Propagation Direction}\label{wave-angle}
The CoMP Doppler velocity image sequence shows coherent fluctuations propagating through
the corona, which are interpreted as propagating kink waves. We begin by determining the 
direction of the propagation of the waves. The coherence between the velocity time-series of each pixel and its neighboring pixels is calculated using an FFT-based method  
(\citealp{MCIetal2008, tom2009}) and correlation maps are derived. Selecting pixels in 
the neighborhood where the coherence value is greater than 0.5 defines a coherence island. This 
coherence island has a distinct direction following the apparent trajectory of the propagating 
waves. The direction of wave propagation is then taken to be aligned with the island, determined 
by fitting a line that minimizes the sum of perpendicular distances from the points to the line. 
This is performed for each pixel in the field-of-view enabling us to create a wave angle map, 
which is displayed in the right panel of Figure~\ref{fig:dopplwave}. The shown angle gives the 
direction of propagation measured counterclockwise from a due East direction. Given that the kink 
mode propagates along the magnetic field, this angle should also represent the magnetic field 
orientation in the plane-of-sky (POS), and this method does indeed show excellent agreement with 
polarimetric measurements of the POS direction of the magnetic field \citep{tom2009}.

\subsection{Determining Wave Power} \label{k-w_diag}
The wave angle map is then used to determine the path of the wave propagation through the 
corona, enabling the kink wave packets to be followed and to determine how they evolve as they
propagate. We select five different wave paths with increasing lengths (center panel of 
Figure~\ref{fig:dopplwave}), where the selected paths
are assumed to follow the quiescent coronal loops and, to satisfy the restrictions of Eq.~\ref{eqn:power_rat}, assumed to represent half the total loop length (this assumption is discussed further in Section~\ref{s:result}). The velocity signal along the wave paths is extracted to 
create time-distance maps, where cubic interpolation is used to map the velocities from 
the selected wave paths onto $(x,t)$ space\footnote{We note that due to the relatively coarse spatial resolution of CoMP, and because the coronal plasma is optically thin, each wave path likely represents the integration over multiple individual loop structures (\citealp{DEMPAS2012}; \citealp{MCIDEP2012}).} For each wave path shown in Figure~\ref{fig:dopplwave}, we also extract the neighboring five wave paths on either side of the original wave path. Each additional path is calculated using the normal vector to the original, and are separated by one pixel in the perpendicular direction. 

\begin{figure*}[!ht]
\centering
\includegraphics[scale=0.75,height=0.98\textwidth,clip]{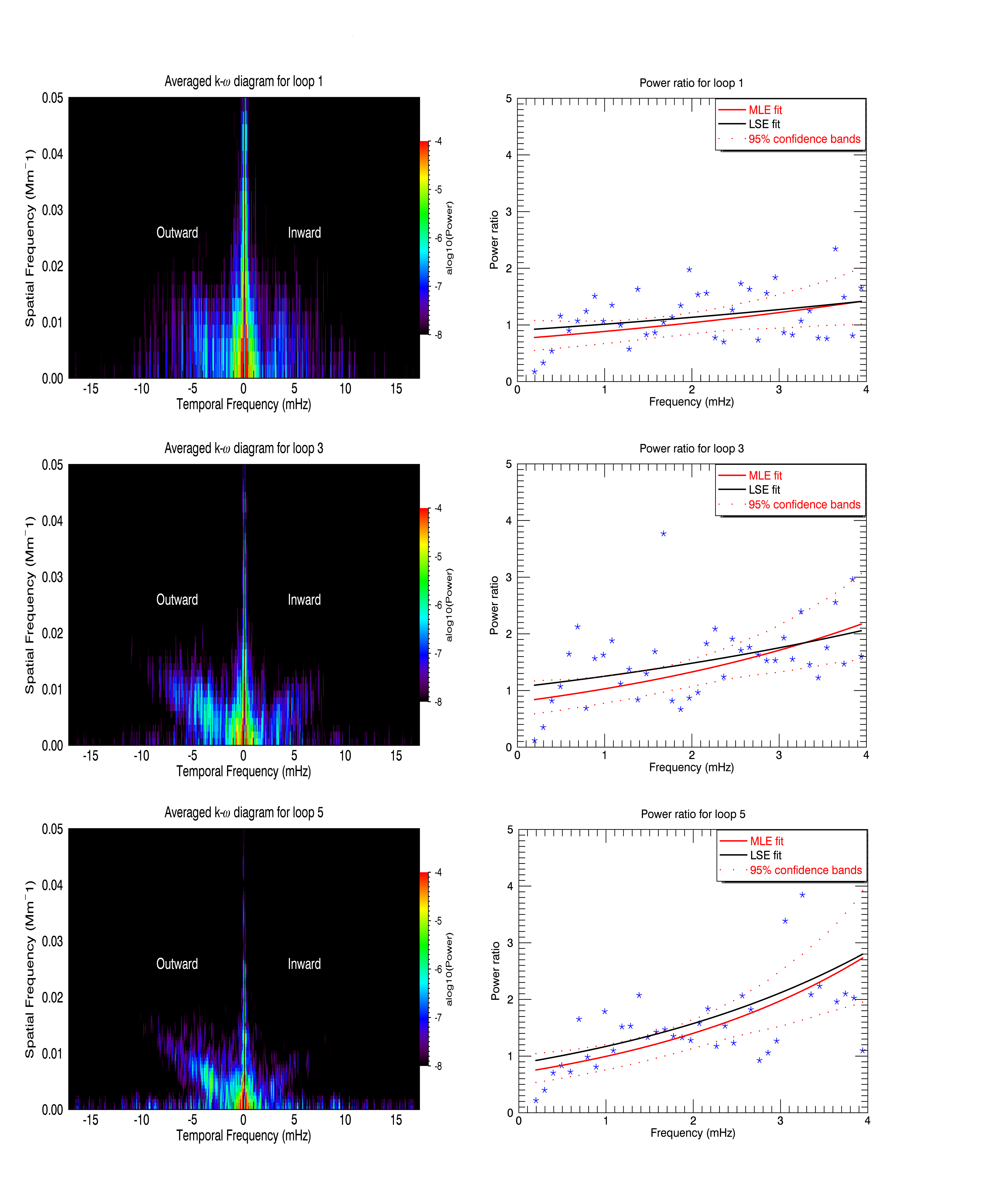}
\caption{Averaged $k-\omega$ diagrams for three selected tracks as shown in Figure \ref{fig:dopplwave}(center). \textit{Top} is 100 Mm; \textit{Middle} is 326 Mm; \textit{Bottom} is 552.6 Mm. The \textit{left} column we show the averaged $k-\omega$ diagrams. The \textit{right} column show the fitted power ratio. The measured power ratio for three coronal loops is shown here by the blue stars, for loops with increasing length. The results from the MLE fitting of the resonant absorption model are over-plotted (red solid), with point-wise Wald confidence bands shown at 95\% (red dotted). As a comparison, the results of the model fit using least-squares (black) is also shown (solid black). }

\label{fig:w-k}
\end{figure*}

These velocity time-distance maps are composed of both the inward
and outward propagating kink waves. Taking a Fourier transform of the velocity time-distance 
maps enables us to produce the $k-\omega$ spectra for velocity power, shown in 
Figure~\ref{fig:w-k}. The wave power is separated for the inward and outward components of the 
wave propagation, and it is evident from all $k-\omega$ spectra that the outward wave power dominates over the inward wave power. By taking the inverse Fourier transform of the inward and outward halves of the 
$k-\omega$ spectra separately \citep{tom2007,tom2009,MORetal2015}, filtered time-distance 
diagrams are created. The filtered time-series are used to obtain the wave propagation speed along the wave 
path for both outward and inward propagating waves. The time-series at the center of the wave 
path are 
cross-correlated to the neighboring time-series along the path. The lag of the 
cross-correlation is determined by fitting a parabola to the peak of the correlation function. 
The propagation speed is then calculated by fitting the slope of the observed lags as a function of the position along the wave path. 

\medskip

Finally, the wave power as a function of frequency for the inward and outward components is 
calculated by summing the spectra in the $k$-direction. For each loop, the inward and outward 
spectra are averaged over the neighboring wave paths in order to suppress the variability
(see Section \ref{subsec:statistics} and Appendix \ref{s:app} for further discussion). From this one dimensional averaged wave power, the ratio of the outward and inward power, $\langle{P(f)}\rangle_{ratio}$, is determined. For each of the 
coronal loops studied, the ratio of power spectra displays an increase in magnitude as frequency increases 
(Figure \ref{fig:w-k}). It is this signature that demonstrates the frequency dependence of the 
change in outward and inward power, indicating that a frequency dependent process is in action 
to attenuate the waves, e.g., resonant absorption. 
  
As discussed in the introduction, in order to estimate the quality factor from the obtained 
power ratio, the model power ratio given by Eq.~(\ref{eqn:power_rat}) should be fit to the data in a robust manner. Therefore, it is necessary to discuss the statistics of the power ratio. 

\subsection{The Statistics of the Power Ratio} \label{subsec:statistics}
In \citet{verth2010} the ratio of the outward and inward spectra were fitted with the model given by Equation (\ref{eqn:power_rat}) using a least-squares minimization. However, as we will show, the assumption that the power ratio values at each frequency ordinate are normally distributed (implicit in least-squares) is incorrect and leads to a poor estimate of model parameters and their uncertainties. Here, we present a new method for the maximum likelihood estimation of model parameters from the ratio of two power spectra obtained via a discrete Fourier transform (DFT).

The power spectra, \textit{I}($f_{i}$) at each frequency ordinate, $f_{i} ;  i={0,1,2,3,...,n}$, from the DFT are distributed about the true power value, \textit{P}($f_{i}$) as

\begin{equation}
I(f_{i})=P(f_{i}) \frac{\chi^{2}_{2}}{2}.
\label{eqn:pow1}
\end{equation}
Here $\chi^{2}_{2}$ represents a random variable from the chi($\chi$)-squared distribution with two degrees of freedom, distributed as 
\begin{equation}
{\chi^{2}_{2}}=\frac{1}{2}\exp\left(-\frac{x}{2}\right)
\end{equation}

 (see e.g., \citealp{vaughan2005}). Suppose now we are interested in taking the ratio of the values $x$ and $y$, drawn from two independent $\chi^{2}_{2}$ distributions $X$ and $Y$. The associated  probability distribution function ($PDF$) is $Z=X/Y$ and the distribution of $Z$ is then given by
\begin{equation}
\psi_{z}=\int_{0}^{\infty}y \psi_{xy}(zy,y)dy.
\end{equation} 
Given that $x$ and $y$ are independent, $\psi_{xy}$ is given by
\begin{equation}
\psi_{xy}=\frac{1}{4}\exp\left({-\frac{x+y}{2}}\right).
\end{equation}
Hence,
\begin{equation} \label{eqn:log-logi}
\psi_{z}=\frac{1}{(1+z)^{2}},
\end{equation}
and the distribution of the ratio of any two given power spectra, $z$ (i.e. ratio of $\chi^{2}_{2}$ distributions) is given by the log-logistic distribution (Eq.~\ref{eqn:log-logi}).
For a non-normalized random variable, $r$, one can obtain the probability distribution by change of variable, introducing 
\begin{equation}
z=\frac{r}{s}
\end{equation}
where $s$ is the appropriate normalizing factor. The resulting $PDF$ is given by
\begin{equation}
g(r)=\psi\left(\frac{r}{s}\right)\frac{dz}{dr}.
\end{equation}
Hence,
\begin{equation}
g(r)=\frac{1}{s}\frac{1}{\left(\frac{r}{s} + 1\right)^{2}}.
\end{equation}
 For the power spectra ordinates, we known that $2I/P$ is $\chi^{2}_{2}$. Hence, if $x = 2I_{1}/P_{1}$ and $y = 2I_{2}/P_{2}$ then, 
 $z = {I_{1}P_{2}}/{I_{2}P_{1}},$ and $r = {I_{1}}/{I_{2}}$, $s = {P_{1}}/{P_{2}}$. Thus,  the $PDF$ of the ratio of the power spectra ordinates is calculated to be
\begin{equation}
g(R_{i}) =\frac{1}{S_{i}} \frac{1}{\left(\frac{R_{i}}{S_{i}}+1\right)^{2}},
\end{equation}
where $R_{i}= {I_{1i}}/{I_{2i}}$ , is the power spectra ratio and $S_{i}={P_{1i}}/{P_{2i}}$ is the true ratio of the spectral power. 

\begin{figure}[!t]
    \centering
    \includegraphics[scale=0.53,clip=true,viewport=0 0 450 380]{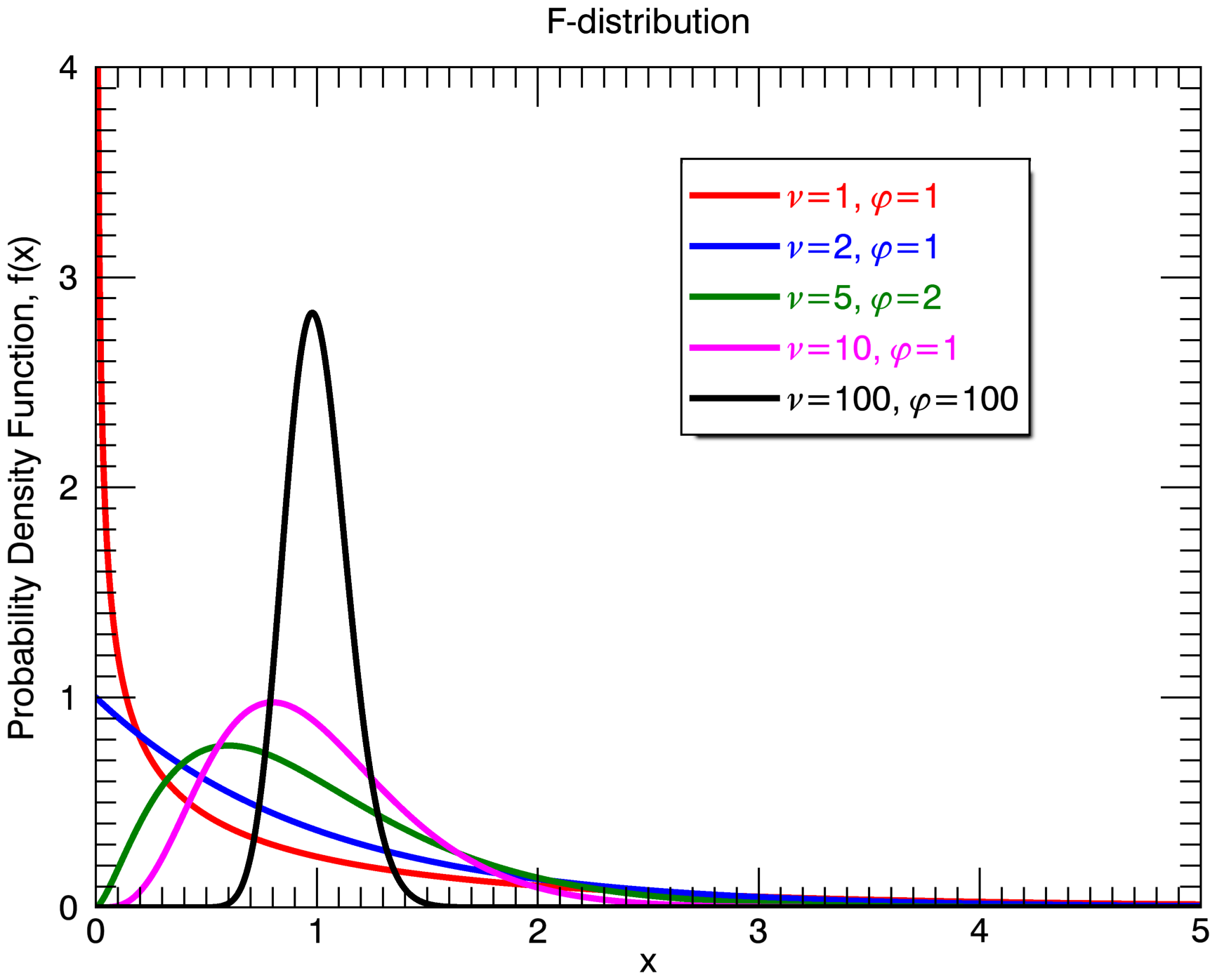}
    \caption{Probability density function for $F$-distribution with different degrees of freedom for a random variable, x. In the case of $\nu$ = $\varphi$ = 100, the density function is log-normally distributed.}
    \label{fig:f-dist}
\end{figure}

\medskip
In this study, several power spectra are summed, which changes the distribution by altering the number of degrees of freedom. 
The ratio of two $\chi^{2}_{\nu}$ distributed variables can be shown to be distributed following the $F$-distribution, given by
\begin{equation}
F(z;\nu,\varphi)= \frac{1}{\beta (\frac{\nu}{2},\frac{\varphi}{2})} {\left(\frac{\nu}{\varphi}\right)}^{\frac{\nu}{2}} {z}^{\frac{\nu}{2} -1} \left(1+z \frac{\nu}{\varphi}\right)^{-\frac{\nu+\varphi}{2}}.  
\end{equation} 
where $\nu$ and $\varphi$ are the degrees of freedom (number of parameters), and $\beta$ is the beta function. 
The log-logistic distribution is recovered for $\nu$ = $\varphi$ = 2. For $\nu = \varphi$, the  $F$-distribution simplifies to
\begin{equation}
F(z;\nu,\nu)= \frac{1}{\beta (\frac{\nu}{2},\frac{\nu}{2})}  {z}^{\frac{\nu}{2} -1} \left(1+z\right)^{-\nu}.  
\end{equation}
The $F$-distribution is an asymmetric distribution with a minimum value 0 and no maximum value. In Figure~\ref{fig:f-dist} we show the nature of the distribution for various values of the degrees of freedom $\nu$ and $\varphi$. There is a different $F$-distribution for each combination of these two degrees of freedom. The distribution is heavily right-skewed for smaller values of $\nu$ and $\varphi$, which means there is a long tail and an increased chance of more extreme large values. As the degrees of freedom increase, the $F$-distribution is more localized.

As before, substituting in the normalized variables gives
\begin{equation}
g_{\nu,\nu}\left(\frac{R_{i}}{S_i}\right) = \frac{1}{S_{i}} \frac{1}{\beta (\frac{\nu}{2},\frac{\nu}{2})}  \left(\frac{R_{i}}{S_{i}}\right) ^{\frac{\nu}{2}-1} \left(1+\frac{R_{i}}{S_{i}}\right)^{-\nu}.
\end{equation} 
Assuming a model $S(\theta)$ for the true power ratio, with unknown parameters $\theta$, the joint probability density of observing $N$ periodogram ratio points $R_{i}$ is given by the likelihood function, $\mathcal{L}$, where

\begin{equation}
\mathcal{L}=\prod^{n}_{i=1} p(R_{i}\vert S_{i})=\prod^{n}_{i=1}\frac{1}{S_{i}}F\left(\frac{R_{i}}{S_{i}};\nu,\nu\right).
\end{equation}

Maximizing the likelihood is equivalent to minimizing the negative of the log of the likelihood function, namely 
\begin{eqnarray}\label{eq:log_lik}
-2\ln{\mathcal{L}}&=& 2\sum_{i=1}^{n}  \left\lbrack \ln S_{i} +\ln \beta\left({\frac{\nu}{2},\frac{\nu}{2}}\right) \right.
\nonumber \\
&+&  \left. \left(1-\frac{\nu}{2}\right) \ln{\left(\frac{R_{i}}{S_{i}}\right)}+ \nu\ln{\left(1+\frac{R_{i}}{S_{i}}\right)} \right\rbrack.
\end{eqnarray}
We have verified that this likelihood function provides consistent estimators for the model parameters $\theta$ (see Appendix \ref{s:app}).

\begin{table*}[!t]

\centering
\begin{tabular}{c c c c c c c} 
\hline\hline 
Loop No. & Half Loop Length   & Power ratio  & $\xi$ & Power ratio  & $\xi$   & Propagation speed \\[0.5ex]
 & (Mm) & (MLE fit) & (MLE fit) &  (least-squares fit) & (least-squares fit) & (km~s$^{-1}$) \\
\hline 
1   & $100\pm7$  & 0.75$\pm$0.17 & 1.83$\pm$0.76 & 0.90 & 2.57 & 687$\pm$17   \\

2   & $197\pm9$  & 0.67$\pm$0.15 & 2.65$\pm$0.55 & 0.91 & 4.46 & 679$\pm$14  \\

3   & $327\pm12$  & 0.80$\pm$0.18 & 3.86$\pm$0.47 & 1.05 & 5.81 & 666$\pm$14   \\

4   & 488$\pm$17  & 0.63$\pm$0.14 & 4.82$\pm$0.43 & 0.78 & 6.17 & 658$\pm$38   \\

5   & 553$\pm$18  & 0.71$\pm$0.16 & 4.76$\pm$0.37 & 0.87 & 5.52 & 676$\pm$15 \\
\hline \hline

\end{tabular}
\caption{Measured loop parameters and wave parameters obtained from MLE. The uncertainties shown correspond to the standard deviation of the mean for MLE parameters and standard deviations for loop parameters. The error in loop length corresponds to the pixel uncertainty of the instrument, the PFSS extrapolation provides us with another uncertainty namely a projection of up to $20^{\circ}$ (0.94 Mm). The least-squares estimates do not have an associated error as we did not have the error estimates for the associated parameters with least-squares fitting.}\label{table1}%
\end{table*}

\subsection{Maximum Likelihood Estimation}\label{s:MLE}
The observed power ratios shown in Figure~\ref{fig:w-k} are then used to estimate the 
model parameters for the power ratio, i.e. the power ratio scaling factor, 
$P_{out}/P_{in}$ and the factor in the exponential, $2L/v_{ph}\xi$ given in Eq.~(\ref{eqn:power_rat}). We use the Powell method for minimization making use of the IDL 
\textit{POWELL} function (e.g., \citealp{BARVAU2012}).

The associated confidence intervals on the model parameters can be estimated by utilizing the \textit{Fisher Matrix} 
$(\mathcal{F})$. The components of $\mathcal{F}_{ij}$ are defined as the expected value of the \textit{Hessian} $(\mathcal{H})$

\begin{equation}\label{eq:fish}
    F_{ij}=\left\langle -\frac{\partial^2 \ln \mathcal{L}}{\partial\theta_i\partial\theta_j}\right\rangle,
\end{equation}
where $\theta$ represents the model parameters \citep{pawitan2001, bevington}. The \textit{Fisher matrix} is a $N \times N$ matrix for $N$ model parameters. The inverse of the \textit{Fisher Matrix} gives the 
covariance matrix, the diagonal elements of which give the standard error squared on each model parameter, $\sigma^2$. The off-diagonal matrix elements provide the covariances between parameters. The \textit{Fisher Matrix} only gives reliable uncertainties when the likelihood surface can be approximated by a multi-dimensional Gaussian. Here we will give the values obtained from the covariance matrix as the estimated parameter uncertainties, and have checked that they are in close agreement with more involved methods of calculating confidence levels, e.g., Wilks confidence intervals \citep{bevington}. At best, the given uncertainties and confidence intervals should be taken as a lower limit. 

We use the standard errors to calculate the point-wise Wald 95\% confidence intervals \citep{bevington} for the model. The likelihood surface and covariance matrix suggest covariance between the model parameters and this is included in the confidence interval calculation. For the measured power ratios given in Figure~\ref{fig:w-k}, the likelihood surfaces are close to a bivariate Gaussian, thus the corresponding confidence bands calculated are reliable. We note that in the case of the ratio of two single (i.e., non-averaged) power spectra ($\nu$ = 2), the likelihood function is irregular and the \textit{Fisher Matrix} will likely provide a poor coverage of the confidence intervals (see examples in Appendix \ref{s:app}). 

\section{Results and Discussion}{\label{s:result}}
\subsection{Potential Field Extrapolation} 
Potential field extrapolations are undertaken with PFSS to determine the geometry for the quiescent coronal loop system shown in Fig.~\ref{fig:dopplwave}. In particular we are keen to examine whether the wave paths determined from following the Alfv\'enic fluctuations are situated in the POS, which has been the implicit assumption in previous analyses (\citealp{tom2009}; \citealp{verth2010}). This assumption has an impact on the measured propagation speeds and lengths of loops, both of which are important quantities for determining the equilibrium parameter $\xi$ from the data (see Equation~\ref{eqn:power_rat}). We note that the plotted magnetic field lines in Figures~\ref{fig:dopplwave} and \ref{fig:pfss} are not supposed to represent the specific coronal loops along which we believe the waves are propagating. However, we expect the extrapolated field to represent the general behavior of the magnetic field in the region, and as such, describe the oscillating loops. Moreover, as mentioned earlier, the spatial resolution of CoMP essentially precludes identifying individual coronal structures. The extrapolated field demonstrates that the loops are approximately situated in the POS, with a maximum angle between the loops and POS found to be 20$^\circ$ (see Figure \ref{fig:pfss}). 

Furthermore, it can be inferred from the extrapolated field lines plotted in the left panel of Figure~\ref{fig:dopplwave} that the geometry of the coronal loops is not symmetric about the apexes. Given that the model used for fitting the wave damping is derived under the assumption that both the outward and inward waves have propagated along half of the loop (Eq.~\ref{eqn:power_rat}), this will likely affect our estimates for $\xi$ (discussed further in Section~\ref{sec:MLAanalysis}). In Appendix~\ref{s:app3}, we give a more general model for the exponential damping that can be fit to the data when measuring over a segment of the loop. Although knowledge of total loop length and the segment length are required, and for this data set there are no stereoscopic data available that would help us achieve this. Moreover, we are hesitant in trying to determine any one-to-one correspondence between the extrapolated field lines and the wave angle guided tracks. Given that the main purpose of this work is to present a more appropriate method for fitting the observed power ratio and demonstrate that the least-squares method gives incorrect model parameters, such a limitation does not invalidate this aim. Hence, the general formula is provided for future work and we ask that these limitations are kept in mind as we proceed with the analysis.

\subsection{Wave Power Analysis}
Using two-dimensional Discrete Fourier transforms, we determine the inward and outward components of 
the wave power corresponding to the Alfv\'enic waves propagating along five wave paths of increasing  
length (wave paths shown in Figure~\ref{fig:dopplwave} center panel). The $k-\omega$ diagrams for three
of the wave paths are displayed in Figure~\ref{fig:w-k} (left column) and provide an indication of 
relative strength of the outward and inward propagating Alfv\'enic waves in the segment of loop under 
consideration. The $k-\omega$ diagrams have the distinct ridges reported in previous observations, 
corresponding to the near dispersion-less kink mode, where the negative frequencies correspond to 
outward waves and positive are inward waves. Given that the spatial frequency-resolution is lower for 
the shorter loops (Figure~\ref{fig:w-k} top left) compared to the longer loops (Figure~\ref{fig:w-k} 
bottom left), the $k-\omega$ diagrams are less well resolved for the shorter loops. In spite of this, 
it can be noticed that as the length of the loop increases, the relative power in the outward 
propagating Alfv\'enic waves to the inward propagating waves increases. Assuming that the Alfv\'enic 
waves entering the corona at both footpoints of the loops have the same power spectra, then this 
potentially has a trivial explanation: For longer loops, the inward propagating waves will have 
traveled further distances and they should be expected to have been damped to a greater degree, as 
suggested by Eq.~(\ref{eqn:power_rat}). Upon collapsing the spectra in the wave number direction and 
taking the ratio of the outward to inward spectra, the plots in the right columns of 
Figure~\ref{fig:w-k} are obtained. The power ratio shows an apparent upward trend as a function of frequency indicating wave damping,
with the relative magnitudes of the power ratio supporting the visual impression from the $k-\omega$ 
diagrams and indicates greater wave damping for the longer loops. Following \cite{verth2010}, we only 
show the ratio of power spectra up to 4~mHz. This is largely because the signal drops below the noise 
level for the inward propagating waves beyond this frequency and leads to a turnover in the power 
spectra.

\subsection{Maximum Likelihood Analysis}\label{sec:MLAanalysis}
Using the derived likelihood function (Eq.~\ref{eq:log_lik}) we are able to fit the power ratio model (Eq.~\ref{eqn:pow1}) to the data points shown in Figure~\ref{fig:w-k}. The maximum likelihood model parameters are used to define the model power ratio curve (red solid line right column of Figure \ref{fig:w-k}), with the values in the covariance matrix enabling us to generate the point-wise Wald confidence bands at $95\%$ via bootstrapping. The confidence bands demonstrate that in each case, there is a clear trend in the power ratio as a function of frequency and supports the idea that frequency-dependent wave damping is in action along each wave path \citep{verth2010}. 

Given that previous work has employed the least-squares method for fitting the power ratio model, we also demonstrate the differences between the parameter estimates from least-squares and MLE methods. In Figure~\ref{fig:w-k} (right column) the model curves obtained from the least-squares (black solid line) are over-plotted and demonstrate that they underestimate the amount of damping present, i.e., corresponding to flatter curves, when compared to the MLE method. 

\begin{figure}[!ht]
\centering
\includegraphics[scale=0.45,trim={0cm 0cm 0cm 0cm},clip]{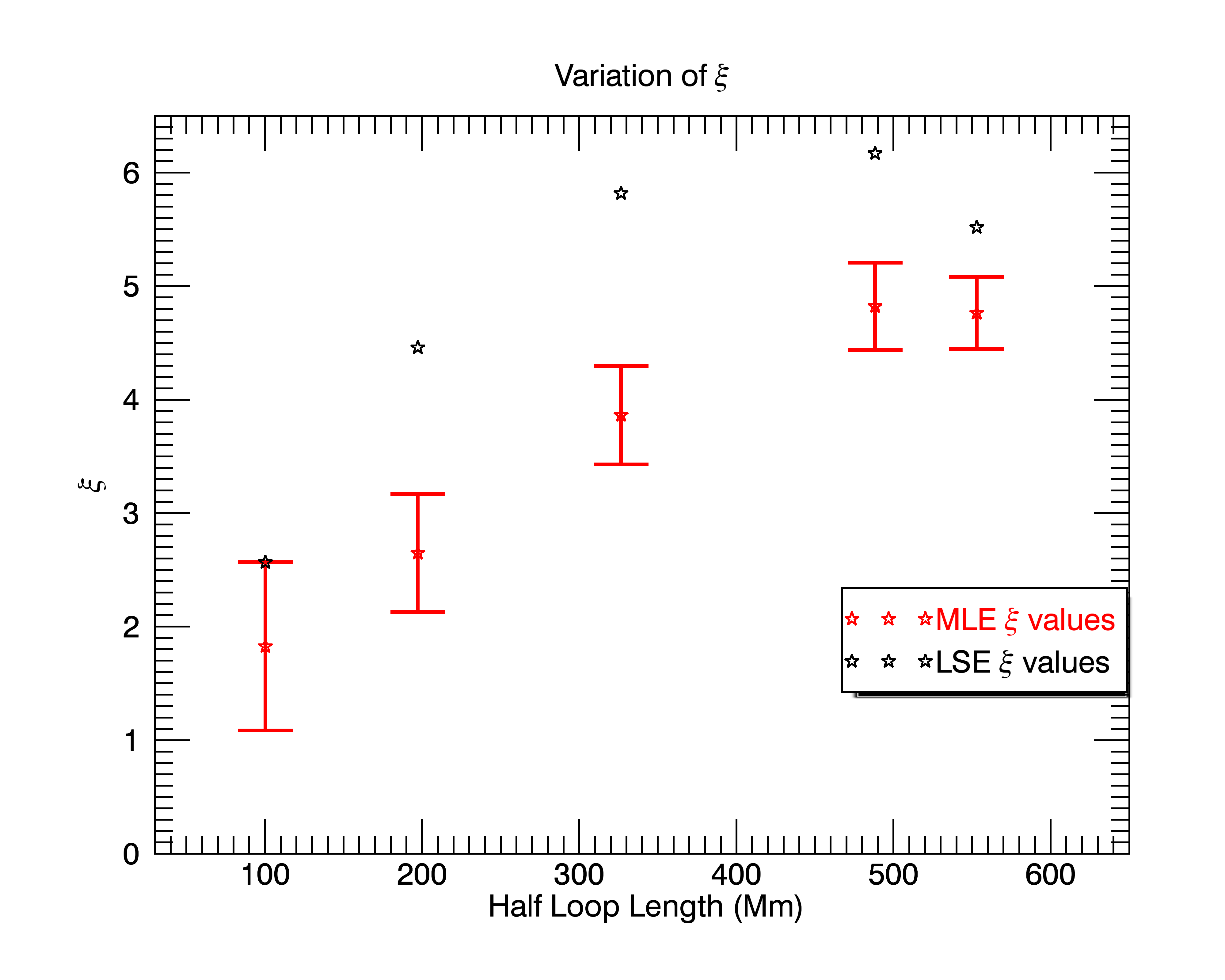}
\caption{Variation of equilibrium parameter $\xi$ with loop length, with associated error bars. The longer loops have higher value of $\xi$.}
\label{fig:xi_var}
\end{figure}

\medskip
To estimate the equilibrium parameters (quality factor), $\xi$, for each selected wave path, we also require the length of the wave path used, assumed to be the half loop length ($L$), and the propagation speed of the Alfv\'enic waves ($\upsilon_{ph}$). The values are summarized in Table~\ref{table1}. The measured propagation speeds of $\approx$680 km {s}$^{-1}$ are consistent with the values obtained in previous studies \citep{tom2009,MORetal2015}. This value is averaged over the outward and inward wave propagation speeds, as we are not considering the potential influence on damping from flows along the loop. The presence of flows leads to modification of the TGV relation \citep{SOLetal2011}.\footnote{We note that any density stratification along the loop will not impact upon the measured power ratio, as any effect on the average amplitude will be the same for both outward and inward waves.} Furthermore, studies of the interaction between the flows (namely the solar wind) and Alfv\'en waves suggest wave action conservation is important, which can result in dissipation-less waves undergoing apparent damping \citep{jacques1977,heinman1980,mckenzie1994,boli2007wind,cranmer2007,chandran2015}. In the case of coronal loops estimating flows is not a trivial endeavor; although the corona is likely to be in a state of thermal non-equilibrium and flows are expected to be present throughout. Several studies have tried to quantify the flow speeds, largely in active region loops, which are typically of the order of 10-50 km {s}$^{-1}$ \citep{reale2010}. Moreover, speeds of 74-123 km {s}$^{-1}$ have also been found in a single event \citep{ofman2008}. These studies suggest the axial flow speed is potentially small compared to the local Alfv\'en speed, and thus we expect this would have little effect on our results. However, further examination of flows in coronal loops is clearly required to assess their impact.

The increase in loop length between the wave paths corresponds to loops reaching higher altitudes in the corona. In Figure~\ref{fig:xi_var}, we show the measured values of $\xi$ as a function of loop length, and our measurements suggest that for the longer loops that reach higher up in the corona, the quality factors increases and, hence, the damping length increases, suggesting the Alfv\'enic waves are subject to a reduced rate of damping. This is in contrast to the $k-\omega$ diagrams and power ratios, which show a greater difference between the outward and inward wave power. This is naturally explained by the fact that the inward waves have propagated further along the longer loops and have been damped to a greater degree than those is the shorter loops, despite the apparent reduced rate of damping in the longer loops. This result is present in both the MLE and least-squares fitting, however the least-squares approach tends to overestimate the fitted values of power ratio and equilibrium parameter.

Given the aforementioned problems with this data set, related to identifying whether the selected wave paths 
are truly half the loop length, we are cautious in our interpretation of this variation in quality factor 
with loop length. A physical explanation for the decrease in damping rate can be made in terms of the density
ratio between the internal and external plasmas. If we assume that the coronal loops are subject to similar 
rates of heating, and the rate of chromospheric evaporation is similar, then the average density of the 
longer loops is likely to be less than those of shorter loops. Hence, compared to the ambient plasma the 
density ratio ($\rho_i/\rho_e$) for longer loops is, on average, less than for the shorter loops.  
Eq.~(\ref{eqn:eqparm}) then implies the equilibrium parameter will increase as the density ratio decreases, 
and matches the observed behavior.

Moreover, the fact that we have potentially not measured the wave path along half a loop will change the model that should be fit to the power ratio (Eq.~\ref{eqn:power_rat_mod}) and alter the measured values of the parameters. Considering the magnetic field extrapolation, there is the possibility that we have measured a loop segment less than half the loop length. In such a scenario, the average power over the this shorter segment, compared to a half loop segment, will be greater for outward waves (as the wave amplitudes averaged over have been damped less over this distance) and less for the inward waves (as the wave amplitudes averaged over will have been subject to greater damping). Hence, the power ratio will be artificially enhanced, giving the appearance of greater damping. This would lead to an underestimate of $\xi$ compared to its true value. Hence, the observed effect of increasing $\xi$ with height would be more pronounced.

\medskip
Finally, it is also worth commenting on the measured value of the factor $P_{out}/P_{in}$, which represents the power of the waves input into the corona at each footpoint of the loop. The power ratio obtained is almost equal to unity in the case of least-squares estimation, consistent to the previous study of \citet{verth2010} and was interpreted as the wave power being generated was the same at both the footpoints. In the case of MLE estimation, we obtain that the power ratio is less than unity implying that the wave power generated at the footpoint associated with inward waves is larger than the other. The current level of uncertainties associated with our measurements does not permit us to rule out that the input power is equal at both footpoints. However, it would not be surprising if the magnitude of the wave power is different at both footpoints, given the physical conditions at the wave source region are likely to be dissimilar.

\section{Conclusion}\label{s:conclusion}
In this paper, we have advanced the methodology for investigating the damping of propagating Alfv\'enic waves from spectroscopic data. The main goal was to provide an improved and more robust method for fitting the ratio of two power spectra, taking into account the statistical properties of the expected distributions of the power ratio for each frequency ordinate. Upon application to a previously studied CoMP data set, we confirmed the previous conclusions that the Alfv\'enic waves are subject to damping, with resonant absorption suspected as main damping mechanism. However, we find that the previously used methodology for fitting the power ratio, i.e., least-squares, has the potential to provide bias estimates of the model parameters, namely the quality factors, $\xi$, and footpoint power ratio $P_{out}/P_{in}$. Importantly, the least-squares fit likely overestimates $\xi$, leading to an underestimation of the strength of the wave damping. An accurate estimate of the quality factor is key in quantifying the rate of energy transfer and the amount of wave energy that might be contributing to plasma heating.

In spite of issues with determining the true geometry of loops in this study, by looking at different wave paths in the data we have been able to find the first potential piece of evidence that the damping length increases as the loop length of loops that reach higher up in the corona increases. The result appears consistent with the result obtained in the case of damped, standing kink waves, where the damping time increases as the loop length increases \citep{verwichte2013}. While it is unclear what may be the underlying cause of this, it could potentially be explained by a decreasing average density ratio between the loop and ambient plasma as loop length increases.

Given the ubiquity of propagating kink waves in the corona has been established, there is a clear need to accurately estimate the damping of propagating kink waves in order to understand the transfer of energy and the contribution of Alfv\'enic wave energy towards plasma heating. The results presented here highlight the need to further investigate the damping of coronal kink waves and provide a robust methodology to achieve this. Future studies should aim to overcome some of the shortfalls associated with the current work.

\acknowledgments
A.K.T thanks Northumbria University for support via a University Research Studentship. A.K.T would also like to acknowledge Tom Van Doorsselaere, Norbert Magyar and Marcel Goossens for valuable discussions. The authors acknowledge the Science and Technology Facilities Council via grant number ST/L006243/1 and for IDL support. The authors acknowledge the work of the National Center for Atmospheric Research/High Altitude Observatory CoMP instrument team.  This work also benefited from discussions at ISSI, Bern (Towards Dynamic Solar Atmospheric Magneto-Seismology with New Generation Instrumentation).

%



\pagebreak

\appendix

\begin{figure*}[!t]
\centering
\includegraphics[scale=0.8,clip=true,viewport=0cm 0cm 22cm 10cm]{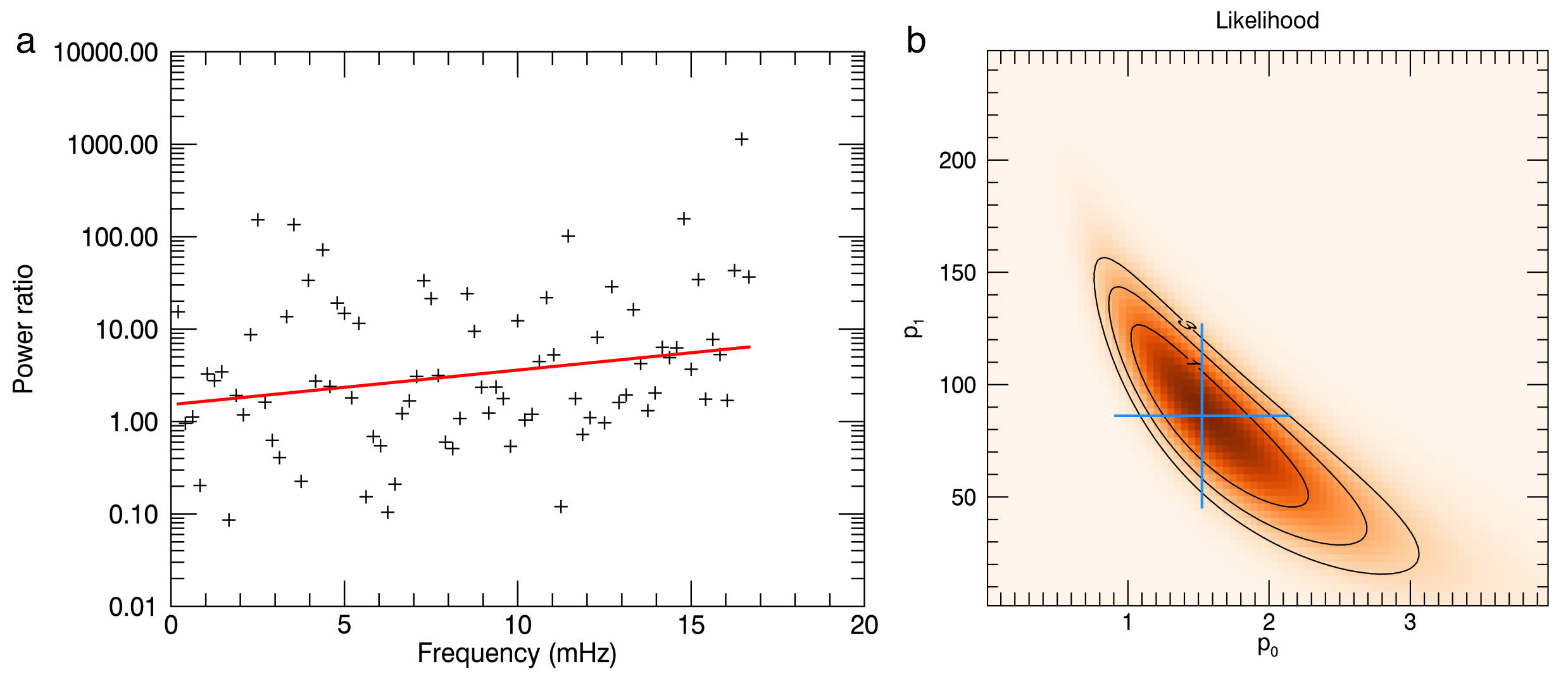}

\caption{Measurement of a synthetic power ratio. a) displays the values of the power ratio as a function of frequency (crosses) with $\nu=2$. The overplotted red line is the MLE estimate for the power ratio curve for the two parameter model. b) is the likelihood surface for the power ratio data, with the color-scale representing the deviance (darker denotes smaller values and lighter denotes larger values). The contours highlight where the deviance of the likelihood surface has values 1, 2, and 3 respectively. The blue lines represent the maximum likelihood estimates of the model parameters. Where the lines cross corresponds to the MLE parameter estimates, with the length of the lines representing the standard error obtained from the covariance matrix. \label{fig:mle_app}}
\end{figure*}

\section{Properties of the Likelihood function}\label{s:app}

Typically, statistical estimators have associated uncertainties composed of the variance and the bias of the estimator, both of which influence the returned value of a model's parameters. In order to demonstrate the suitability of our derived likelihood function (Eq.~\ref{eq:log_lik}) and its performance for measuring model parameters, we perform Monte Carlo simulations that mimic the MLE fitting process discussed in Section~\ref{s:MLE}. 

\medskip

Random synthetic time-series are generated following the method suggested by \cite{TIMKOE1995}, where the power spectra of the time-series is designed to match typical values from observed coronal power spectra \citep[][]{MORetal2016, MORetal2019}. The time-series are produced in pairs, one representing the outward waves and the other series having a power spectra multiplied by a frequency dependent exponential term to represent the damping of the inward waves. The ratio of the power-spectra for the time-series is calculated following the same methodology described in the main text. A model of the form $p_0\exp(p_1f)$ is fit to the ratio of the power spectra using maximum likelihood, where the true values are $p_0=1$ and $p_1=100$. To demonstrate the expected behavior of the power ratio, the MLE parameter estimation and the likelihood surface, we present a couple of illustrative examples. Synthetic time-series are calculated for a fixed length of 160 data points. 

First, it is worth examining what happens when the ratio of two power spectra are taken with no averaging, i.e., $\nu=2$. Figure~\ref{fig:mle_app}a shows the values of the ratio of two power spectra. Although the shape of the true power spectra, and hence the ratio, are smooth functions, the inherent distribution of the power spectra ordinates can occasionally lead to some extreme large values when the ratio is taken. For example, power ratio values of $\approx100$ and $\approx1000$ are obtained, even though the actual value of the underlying power ratio never exceeds 10 for the given frequency range. This is of course a reflection on the skewed nature of the $F$-distribution. It also highlights the potential for a least-squares fitting of the power ratio to provide inaccurate parameter estimates, as it will tend to provide parameter values that balance the number of points that fall on either side of the model power ratio.

The MLE estimated parameters provide model shown by the red line, which is a reasonable match to the true ratio curve. The corresponding likelihood surface is shown in Figure~\ref{fig:mle_app}b, and can be observed to be irregular, i.e., highly non-Gaussian. The model parameters also show clear evidence for covariance. The contours plot the isocurves of the deviance, which can be used to define confidence intervals for parameters \citep{pawitan2001}. For a two parameter model, the 68\% confidence intervals for parameters can be estimated from the likelihood ratio method. This corresponds to parameter values in the region of the likelihood surface with a deviance of 2.27 or less. In Figure~\ref{fig:mle_app}, we also show the standard errors on the model parameters obtained from the covariance matrix (the inverse of Eq.~\ref{eq:fish}).
While in this case the standard errors do contain the true values of the model parameters, they under-estimate and misrepresent the uncertainty, which is asymmetric about the estimated parameters due to the irregular shape of the likelihood surface.

\begin{figure}[!t]
\centering
\includegraphics[scale=0.8,clip=true,viewport=0cm 0cm 22cm 9cm]{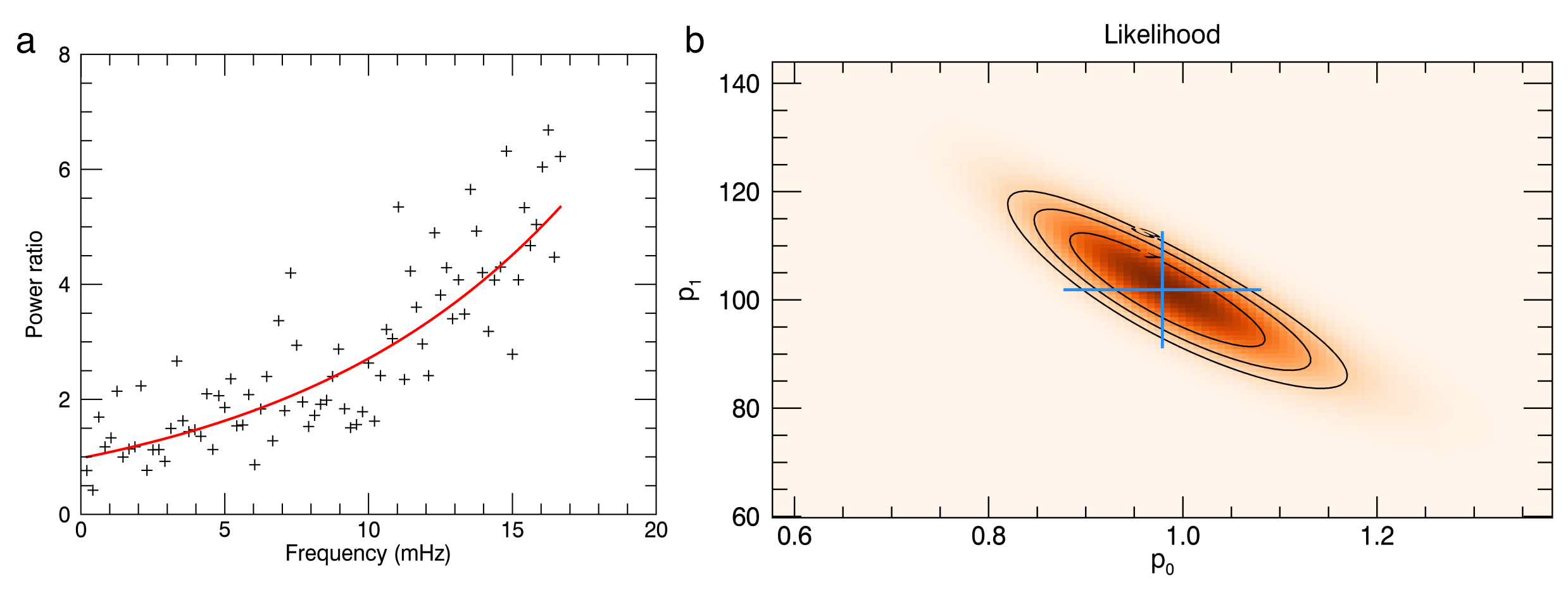}

\caption{Same as Figure~\ref{fig:mle_app}. Here the results show the ratio of the average of 20 power spectra ($\nu=40$). \label{fig:mle_app2}}
\end{figure}

The above discussed example is an extreme case and in reality, it is typically possible to average together a number of 
neighboring time-series, or if the series is long enough it can be segmented into multiple, shorter time-series. Let us now examine a 
case where 20 outward and 20 inward spectra are averaged together before taking their ratio. The results are noticeably different and 
in Figure~\ref{fig:mle_app2} the `measured' values of the power ratio do not possess such extreme deviations from the true values. The 
corresponding likelihood surface from the parameter estimation is more regular, i.e., 
closer to a two-dimensional Gaussian. It can also be noticed that the standard errors from the covariance matrix provide a better representation of the uncertainty.

\smallskip

\begin{figure}[!b]
    \centering
    \includegraphics[scale=0.8,clip=true,viewport=0 0 600 240]{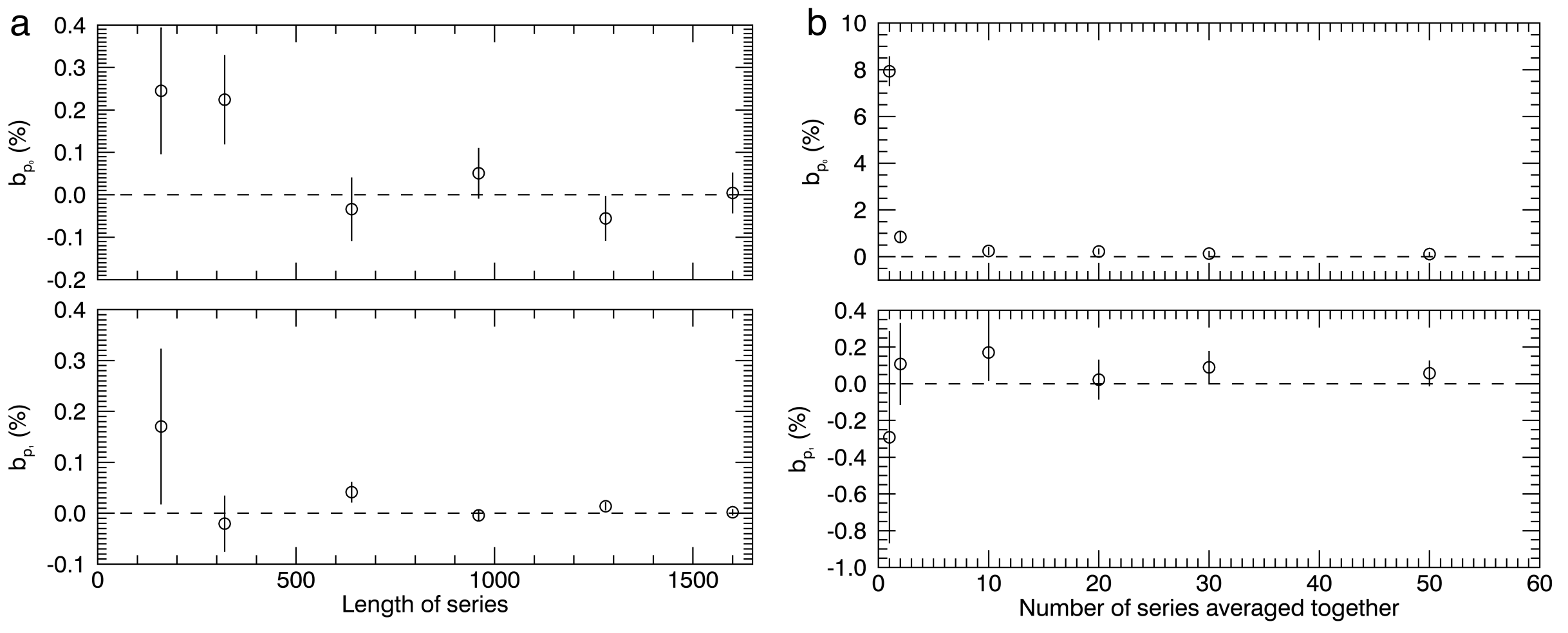}
    \caption{Measured biases on model parameters ($p_0$, $p_1$) from Maximum Likelihood Estimation. The biases are derived from the mean value of the differences between the input parameter value and the measured value form the MLE, expressed as percentages. The error bars show the standard error of the biases calculated from the repetitions of the simulations. }
    \label{fig:sim_res}
\end{figure}

Now, we aim to demonstrate the properties of the bias on the model parameters, where the bias is the difference between the measured value and the true value. We create various sets of simulations that are composed of 5000 repetitions each. We choose to modify two properties of the measurement process in order to demonstrate the influence on the accuracy of the MLE. First, the frequency resolution of the signal is changed by increasing the length of the time-series, which leads to more data points being available for the model fitting. This would mimic the behavior of extended observations at the same cadence, or increasing the observational cadence. For this set of simulations, we choose to average over 10 pairs of power-spectra before taking their ratio ($\nu=20$). In Figure~\ref{fig:sim_res}a we show the value of the bias on the MLE parameters as a percentage and how this varies with signal length. The shown bias is the average value over the 5000 repetitions and the associated standard error on the bias. It can be seen that the MLE estimates for the parameters are asymptotically unbiased, i.e., tend to the correct value as the number of data points included in the fit increases. However, for shorter length data sets there may be some bias in the measured result, although this is somewhat negligible depending upon the variance of parameter estimates.  

The second set of simulations varies the number of power spectra that are averaged together, i.e., varying the degrees of freedom $\nu$, keeping the length of the initial time-series as 160 data points. The results are shown in Figure~\ref{fig:sim_res}b. The plots demonstrate that the number of power spectra averaged together for the ratio has a dramatic affect on the bias of the MLE. If only a single power spectra is used for outward waves and also for inward waves ($\nu=2$), then the bias in the $p_0$ parameter is as much as 8\%. However, the bias significantly decreases when two or more power spectra are averaged together and may be considered negligible depending upon the variance of the model parameters. 

\medskip

\begin{figure}[!t]
\centering
\includegraphics[scale=0.55]{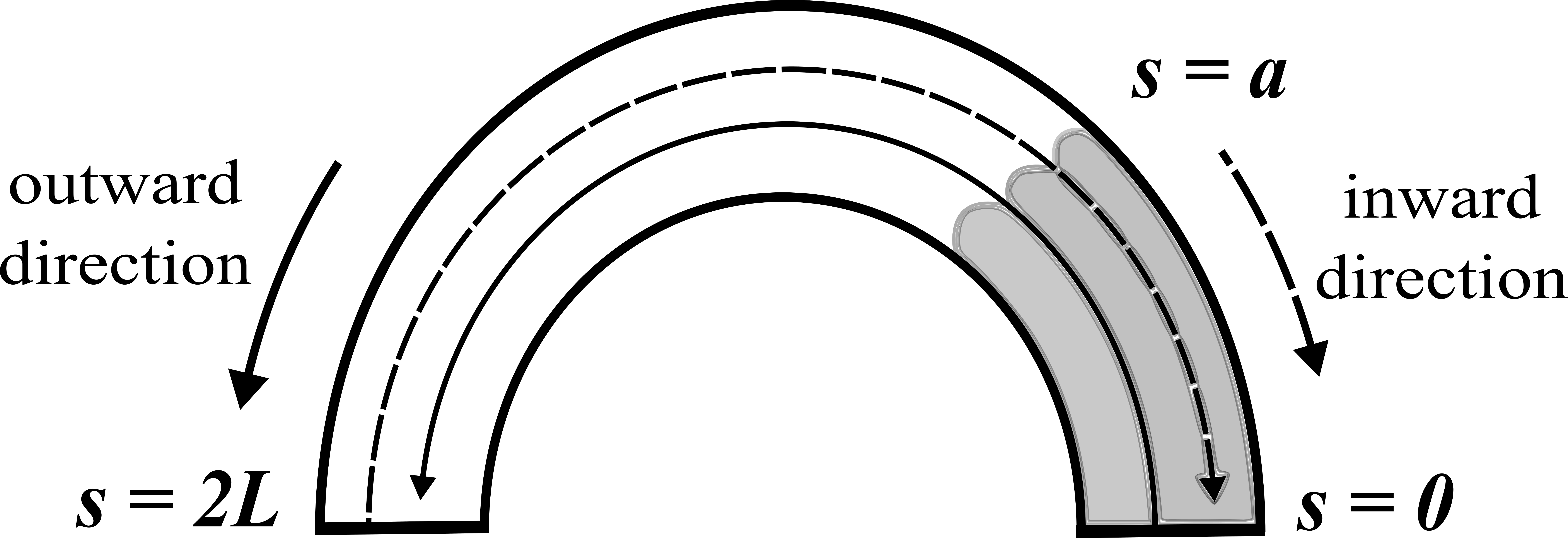}
\caption{Simple illustration of the observed semi-circular geometry of the coronal loop system. The direction of outward and inward wave propagation is shown by the arrows. \label{fig:model}}
\end{figure}

\section{A Modified Model}\label{s:app3}
Here we derive a model for the power ratio that takes into account that only a segment of the loop can be measured. A schematic of physical situation is shown in Figure \ref{fig:model}, where one is only able to measure wave behavior in the shaded section of the loop, from the footpoint, $s=0$, to $s=a$. The average power over the segment associated with the outward propagating waves is given by 
\begin{equation}
\langle{P(f)}\rangle_{out}= \frac{1}{a}\int_{0}^{a} P_{out}(f) \exp \left(-\frac{2f}{\upsilon_{ph}\xi_{E}} s\right) ds.
\end{equation}

Similarly the average power associated with the inward propagating waves is given by, 
\begin{equation}
\langle{P(f)}\rangle_{in}=\frac{1}{a} \int_{2L-a}^{2L} P_{in}(f) \exp \left(-\frac{2f}{\upsilon_{ph}\xi_{E}} s\right) ds.
\end{equation}
If we take the ratio to obtain the power ratio $\frac{\langle{P(f)}\rangle_{out}}{\langle{P(f)}\rangle_{in}}$ we obtain the expression

\begin{equation}
\langle{P(f)}\rangle_{ratio}=\frac{P_{out}}{P_{in}} \frac{\exp\left({\frac{-2fa}{\upsilon_{ph}\xi}}\right)-1} {\exp\left(\frac{-4Lf}{\upsilon_{ph}\xi}\right)-\exp\left(\frac{-2f(2L-a)}{\upsilon_{ph}\xi}\right)}. \label{eqn:power_rat_mod}
\end{equation}
This equation can then be used as the model for MLE to obtain an estimate of damping length when examining only a segment of the loop, only if a reasonable estimate for $a$ is known.



\pagebreak
\bibliographystyle{aasjournal}
\bibliography{paper_damping} 



\end{document}